\documentclass{optica-article}

\journal{opticajournal} 

\articletype{Research Article}

\usepackage{mathtools, cuted}
\usepackage{color}
\usepackage{multicol,lipsum}
\DeclarePairedDelimiter\bra{\langle}{\rvert}
\DeclarePairedDelimiter\ket{\lvert}{\rangle}
\usepackage[labelfont=bf]{caption}
\usepackage{lineno}

\begin{document}


\title{A Variational Approach to Learning Photonic Unitary Operators}

\author{Hadrian Bezuidenhout \authormark{1}, Mwezi Koni\authormark{1}, Jonathan Leach\authormark{2}, Paola Concha Obando\authormark{1}, Andrew Forbes\authormark{1} and Isaac Nape\authormark{1*}}
\address{\authormark{1} School of Physics, University of the Witwatersrand, Private Bag 3, Wits 2050, South Africa\\
\authormark{2} Heriot-Watt University, Edinburgh Campus, EH14 4AS
}

\email{\authormark{*}Isaac.Nape@wits.ac.za} 

\begin{abstract*} Structured light, light tailored in its internal degrees of freedom, has become topical in numerous quantum and classical information processing protocols. In this work, we harness the high dimensional nature of structured light modulated in the transverse spatial degree of freedom to realise an adaptable scheme for learning unitary operations. Our approach borrows from concepts in variational quantum computing, where a search or optimisation problem is mapped onto the task of finding a minimum ground state energy for a given energy/goal function. We achieve this by a pseudo-random walk procedure over the parameter space of the unitary operation, implemented with optical matrix-vector multiplication enacted on arrays of Gaussian modes by exploiting the partial Fourier transforming capabilities of a cylindrical lens in the transverse degree of freedom for the measurement. We outline the concept theoretically, and experimentally demonstrate that we are able to learn optical unitary matrices for dimensions $d$ = 2, 4, 8 and 16 with average fidelities of $>90\%$. Our work advances high dimensional information processing and can be adapted to both process and quantum state tomography of unknown states and channels.
\end{abstract*}
\section{Introduction}
\label{sec:Intro} 
Structured light is a valuable resource for information processing \cite{forbes2021structured, nape2023quantum} as it offers the opportunity for increasing the encoding capacity of many protocols. In particular, structured light encoded in the transverse spatial degree of freedom, such as its orbital angular momentum \cite{yao2011orbital},  position \cite{da2021path} and  pixel/position \cite{valencia2020unscrambling, valencia2020high}, is finding applications in communications \cite{wang2022orbital, huang2014100},  cryptography \cite{cozzolino2019high, mirhosseini2015high, sit2017high, shikder2022image, wang2021high}, imaging \cite{sun20133d, sephton2023revealing, moodley2023all}, and computing \cite{ cheng2020simulate, perez2018first, kumar2022optical}. However, a crucial task is to characterise the channels  that the photons/laser fields are propagated through, and undoing any perturbations that may have acted on the field\cite{valencia2020unscrambling}. This often involves finding and undoing the transmission matrix, a tedious process that can be costly from both  the measurement and numerical inversion perspectives in both  classical \cite{dharnidharka2021optical} and quantum  \cite{mohseni2008quantum} systems, even if the channel is unitary. Here we ask: can we instead use re-configurable system that can steer itself towards the channel unitary operator by learning it iteratively?\break \\
Remarkably, hybrid quantum-classical quantum algorithms that merge the adaptability of quantum circuits with the robustness of classical optimisation techniques have recently surfaced \cite{endo2021hybrid}, hinting at a answer to this question. Specifically, variational quantum algorithms (VQAs) \cite{cerezo2021variational}  are exemplary of this; VQAs exploit the ability to encode a parameterised unitary operator into a quantum circuit, whose parameters are iteratively adjusted until the solution to a given problem is found.  This framework is implemented in various quantum computing platforms to model the complex dynamics of molecules \cite{peruzzo2014variational}, solving systems of equations in linear algebra \cite{xu2021variational, bravo2023variational},  solving quadratic combinatorial optimisation problems \cite{atchade2022quantum} and for quantum machine learning \cite{martin2022quantum} (see review on VQA's \cite{cerezo2021variational}). By setting up a problem Hamiltonian (goal/energy function), a parameterised unitary can be used to probe possible solutions to a given problem. Accordingly, the parameters are iteratively updated such as to minimise the goal function; which is analogous to finding the ground state energy of a Hamiltonian. A classical computer is then used to check if the goal function is being minimised and adapts the unitary parameters accordingly. Because quantum computers inherently utilise matrix-vector multiplication between operators and states, if the VQA framework is to be used optically, a similar mathematical structure is required for transverse spatial modes and for encoding adaptable unitaries. Notably, digital devices such as  spatial light modulators \cite{forbes2016creation, ayoub2021high} and more recently, re-configurable metasurfaces \cite{zhang2020optically} can be used to encode parameterised unitaries in the form of phase masks; additionally, vector matrix multiplication can be achieved using SLMs and Fourier lenses \cite{spall2020fully}.\break \\ 
 In this article, we blend the parallel processing power of optical vector-matrix multiplication together with variational computing principles to develop a method for learning unitary operations in an optical system. We utilise basis states of lattice Gaussian modes encoded as an array on a grid and exploit the tensor product structure of the Cartesian plane to realize a parallel processing approach to evaluating the vector-matrix multiplication, while traversing the solution space of the desired unitary matrix using a random walk on its parameter space. Our variational approach finds the unitary operator by minimising a goal function that finds its minimum once the target unitary operation is found. We demonstrate this for dimensions d = 2, 4, 8 and 16 as a proof of principle, reaching fidelities above 90\% in all cases.
\section{Theory}
\label{sec:Theory}
\begin{figure}[!t]
    \centering\includegraphics[keepaspectratio, width=1\textwidth]{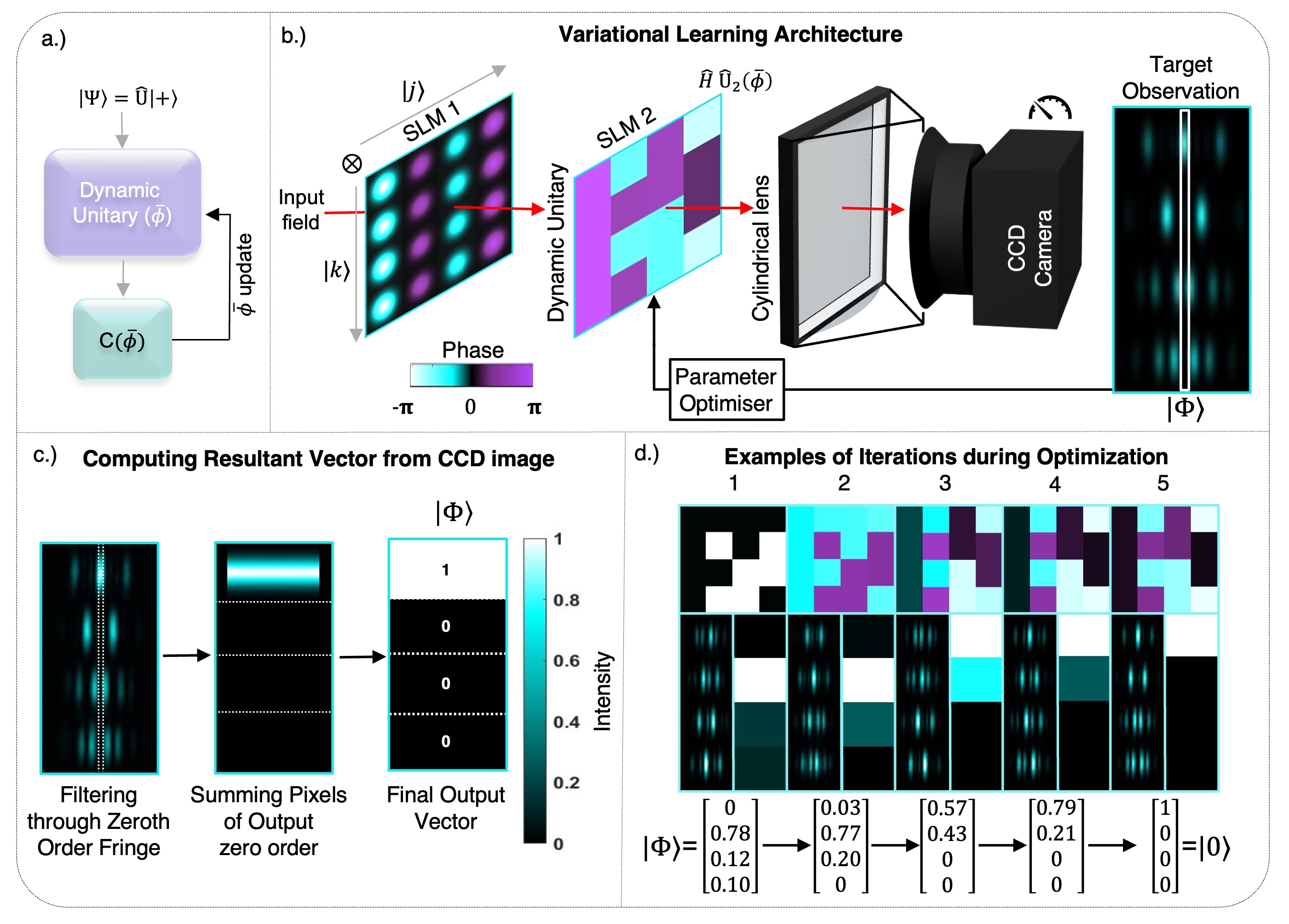}
    \caption{Unitary Optimisation \& Output Processing. (a) Adaptive process for finding an unknown Unitary matrix ($\hat{U}$) by utilising a dynamic (parameterised) unitary matrix that has adaptable parameters $\bar{\phi}$. A state is prepared and acted on with the Unitary $\hat{U}$, thereafter, the dynamic unitary acts on it and the output is measured. The aim is to minimise a  goal function $C(\bar{\phi})$ that depends on the measured output. (b) An input beam with a uniform phase  profile (a planewave)  is encoded with an unknown phase profile  on SLM 1 embedding the state vector, $\ket{\Psi}$, and transferred to SLM 2 where a dynamic unitary ($\hat{H}\hat{U}_2(\bar{\phi})$) is encoded.  The matrix vector multiplication is then realised using a cylindrical lens and the result emerges in the farfield, encoded into an interference pattern.  Upon measurement with the CCD camera, the measured pattern in then used to compute the goal function, after which the parameters of the dynamic unitary are then updated and the processes is repeated. The process repeats until the cost function reaches a minimum, i.e. until the target interference pattern is obtained.  (c) Filtering through the zeroth order mode to determine output vector from the optical matrix vector multiplication. Because the cylindrical lens interferes the Gaussian lattice horizontally, the result is measure in the vertical direction. We only extract the intensity of the filtered components. (d) Columns from left to right show a simulation of the dynamic matrix (top), the observed pattern (bottom left) and the summed filtered intensities (bottom right) during multiple iterations as it converges to the optimum solution (target interference pattern).}
    \label{fig:concept}
\end{figure}
\subsection{Variational Approach to Learning Unitary Matrices}
\label{subsec:Variational Approach}
Our approach is illustrated graphically in Fig.~\ref{fig:concept} (a). Suppose a unitary $\hat{U} \equiv \hat{U}(\bar{\theta})$, with parameters $\bar{\theta}$, that satisfies $\hat{U}\hat{U}^{\dag}=\mathbb{I}$, acts on a uniform superposition $\ket{+} \propto \sum^{n-1}_{j = 0} \ket{j}$ of $n$ states to produce the state
    \begin{equation}
        \begin{split}
            \ket{\Psi} &= \sum_{j=0}^{n-1} a_{j}e^{i \theta_{j}}\ket{j},\\
            &=\sum_{j=0}^{n-1} \Psi_j \ket{j}.
        \end{split} 
        \label{eq: psi}
    \end{equation} 
It is a straight forward exercise to see that by undoing the action of the unitary,  by applying the inverse of the operation $\hat{U}^{-1} = \hat{U}^\dagger$  to  the state $\ket{\Psi}$, we can map the state back to the uniform superposition $\ket{+}$. Furthermore, because $\ket{+} = \hat{H} \ket{0}$,  where $\hat{H}$ is the Hadamard transform satisfying $\mathbb{I} = \hat{H}\hat{H}^\dagger =\hat{H}\hat{H}$, then after applying the operation $\hat{H}\hat{U}^\dagger$ to $\ket{\Psi}$  we obtain the logical basis state $\ket{0} \equiv (1, 0, 0, ...0)^\dagger$.  As shown,  our goal is to find the operator $HU^\dagger$ that maps this arbitrary state $\ket{\Psi}$ back onto the logical basis state $\ket{0}$. This may be crucial for extracting the transmission matrix of a unitary channel that may act on a optical field, e.g., optical fiber \cite{valencia2020unscrambling}, or atmospheric turbulence \cite{ndagano2017characterizing}.\break \\
To achieve this, we construct a parameterised unitary $\hat{U}_2(\bar{\phi})$ - Without loss of generality and to keep the first demonstration intuitive, we set the amplitude coefficients of the input state $\ket{\Psi}$ to $a_{j}=1$ and $\hat{U}$ to be diagonal, but with arbitrary phases. Our parameterised unitary is then given by
    \begin{equation}
    \label{eqn:diagonalunitary}
        \begin{split}
            \hat{U}_2(\bar{\phi}) = 
            \begin{pmatrix}
                e^{i\phi_{1}} & \dots & 0\\
                \vdots & \ddots & \vdots\\
                0 & \dots & e^{i\phi_{n}}   
            \end{pmatrix} \text{ and }
            &\hat{U}_2^{\dagger}(\bar{\phi}) = 
            \begin{pmatrix}
                e^{-i\phi_{1}} & \dots & 0\\
                \vdots & \ddots & \vdots\\
                0 & \dots & e^{-i\phi_{n}}   
            \end{pmatrix}.
        \end{split}
    \end{equation}  
This unitary acts on $\ket{\Psi}$ to produce the state
    \begin{equation}
         \label{eqn:phaselink}
         \hat{U}_{2}^{\dagger}\ket{\Psi} = \sum_{j}e^{i\theta_{j}}e^{-i\phi_{j}}\ket{j} = \sum_{j}e^{i(\theta_{j}-\phi_{j})}\ket{j},
     \end{equation}
where we see that the phases contained within $\bar{\theta}$ are uniquely linked with those of $\bar{\phi}$. This allows us to perform a random walk on the parameters $\bar{\phi}=[\phi_{0},\phi_{1}, \dots ,\phi_{n}]$  until
\begin{align}
  \ket{\Phi} \equiv 
\hat{H}\hat{U}_2(\bar{\phi})\ket{\Psi} &= \sum^{n-1}_j \hat{H}\hat{U}_2(\bar{\phi}) \Psi_j  \ket{j}\\ 
&\rightarrow \ket{0}.
\label{eq: finalState}
\end{align}
To ensure that the operator is found, we minimise the square of the trace distance 
between the states $\ket{\Phi} = \hat{H}\hat{U}_2(\bar{\phi})\ket{\Psi} \equiv \hat{H} \hat{U}_2(\bar{\phi}) \hat{U}\hat{H} \ket{0}$ and $\ket{0}$. 
Given such a problem, a suitable cost function is \cite{cerezo2021cost}
\begin{align}
 C(\bar{\phi}) &= \text{Tr} \left( H_{0}\ket{\Phi}\bra{\Phi} \right),\\
&= 1 - |\bra{0}\hat{H} \hat{U}_2(\bar{\phi}) \ket{\Psi}|^2,
\label{eq: goal}
\end{align}
where 
\begin{equation}
H_{0} = \mathbb{I}  - \ket{0}\bra{0},
\end{equation}
satisfying $H_{0} = H^{\dagger}_{0}$.  As such, this problem can be understood from the perspective of variational quantum algorithms (VQA) \cite{cerezo2021variational, bravo2023variational} where our system aims to minimise $C(\bar{\phi})$ using a parameterised unitary,  $\hat{U}_2(\bar{\phi})$, to search the solution space, while a classical computer is used to select parameters in the direction of steepest descent. In this way, searching for the optimal parameters $\bar{\phi}$, is similar to finding the minimum eigenvalue of  $H_{0}$; this is the main goal of VQAs. Once  $\bar{\phi} = \bar{\phi}_0$ is found, then we are guaranteed that $\hat{U}_2(\bar\phi) = \hat{U}^\dagger(\bar{\theta})$ since this minimises our goal function.

\subsection{Optical implementation}
\label{subsec:opticalimplementation}
We first outline conceptually how this might be implemented optically before outlining the experimental realisation in the section to follow. To implement our approach, we make use of a fully re-configurable matrix multiplication system \cite{spall2020fully}, consisting of spatial light modulators and a cylindrical lens shown in Fig. \ref{fig:concept}(b). We can define transverse spatial modes of light using discrete position basis states, $\ket{j}\ket{k} \equiv \ket{x_j}\ket{y_k} \in \mathbb{R}_2$, tensor products of discrete positions of states defined in the $x$ and $y$ Cartesian directions. As such we construct basis elements composed of a lattice/grid of non-overlapping Gaussian modes, with each state on the lattice given by
\begin{equation}
    \ket{j}\ket{k} = \int G(\bar{r} - \bar{r}_{jk} ) d^2r \ket{x}\ket{y} 
\end{equation}
where $\bar{r} = (x, y) \in \mathbb{R}_2$ are the Cartesian coordinates while $\bar{r}_{jk} = ( x_{j}, y_{k})$ are the centres of each Gaussian mode, and $G(\bar{r} - \bar{r}_{jk} ) \propto e^{-|\bar{r} - \bar{r}_{jk}|^2 / w_0^2}$ are the displaced Gaussian mode functions with $w_0$ representing the waist beam radius of each mode.\\ \\ The optical operation is as follows: an input plane wave with uniform intensity and phase, as shown by the input beam in Fig. \ref{fig:concept}(b), is initialised by the first spatial light modulator (SLM 1) that encodes the desired Gaussian lattice
\begin{equation}
    \label{eqn:emptystate}
        \ket{+}\ket{+}=\sum_{jk}\ket{j}\ket{k}.  
\end{equation}
The same SLM simultaneously implements the operator $\hat{V} = \hat{U} \otimes \mathbb{I}$ which encodes the state $\ket{\Psi}$ from Eq.~(\ref{eq: psi}) across the elements in the $x$-direction while repeating them in the $y$-direction \cite{spall2020fully}, resulting in the state
    \begin{align}    
\ket{+}\ket{+} \xrightarrow{\hat{V}} &= \sum_{jk} \hat{U} \otimes \mathbb{I} \ket{j}\ket{k} \\ &= \sum_{jk} \Psi_j \ket{j}\ket{k}\\
&= \ket{\Psi}\ket{+}.
\label{eq: latticestate}
\end{align}
 Here the $x$-coordinate encodes the state while the $y$-coordinate will be used later for extracting the optical vector matrix multiplication result.  The state in Eq~(\ref{eq: latticestate}) is then acted on by the dynamic operator $\hat{M} =  \hat{H}\hat{U}_2(\bar{\phi})$ at the second spatial light modulator (SLM 2), which encodes the combination of a Hadamard matrix $\hat{H}$ and our dynamic unitary $\hat{U}_2(\bar{\phi})$, represented by
     \begin{align}
\hat{M} (\bar{\phi})  \ket{\Psi}\ket{+} &= \sum_{k=0}^{n-1} \hat{M}_{kj} (\bar{\phi}) \ket{\Psi}\ket{k}, \\
 &= \sum^{n-1}_{j=0} \hat{M}_{kj} (\bar{\phi}) \Psi_j \ket{j}\ket{k}.
 \label{Eq: statebeforelens}
\end{align}
Propagating the corresponding field through a cylindrical lens applies a one-dimensional Fourier transform in the $x$-coordinate, which is analogous to a summing over the elements of each row of the matrix. Subsequently, by measuring the central zero order components in the $y$-direction we obtain a column vector representing the desired multiplication in the $y$-direction. This is akin to integrating the $j$ components in the first ket ($\ket{j}_x\ket{k}_y$), therefore, we can extract the resulting vector elements in the second ket with a single shot measurement, i.e., with a CCD camera from an array of pixels as in Fig.~\ref{fig:concept}(c), shown here for the case when the resulting product produces the $\ket{\Phi} = \ket{0}$ state as an output while other examples are shown Fig. \ref{fig:concept} (d).   Accordingly, this leads to a result similar to Eq.~(\ref{Eq: statebeforelens}) and so the state we obtain is
\begin{align}
  \mathcal{N} \sum^{n-1}_{k=0} (\sum^{n-1}_{j=0}  \left( \hat{M}_{kj} (\bar{\phi})  \Psi_{j} \right) \ket{k} = \mathcal{N} \sum^{n-1}_{k=0} \Phi_k  \ket{k},
\end{align}
where $\mathcal{N}$ is the normalisation factor and  $\Phi_k$ are the components of $\ket{\Phi} =  \hat{M}(\bar{\phi})\ket{\Psi}$  for $\hat{M}(\bar{\phi}) = \hat{H}\hat{U}_2(\bar{\phi})$ as desired (see Eq.~(\ref{eq: finalState})).  After measuring the norm squared of the vector components, $\Phi_k$, from the column vector result of the optical vector matrix multiplication, we can compute the goal function 
\begin{equation}
    C(\bar{\phi}) = \sum^{n-1}_{k=0} | |\Phi_k|^2  - \delta_{k0}|^2,
\end{equation}
where $\delta_{kl}$ is the Kronecker delta and $\delta_{k0} =\langle k| 0 \rangle$. The equation above measures the trace distance squared between the state $\ket{\Phi}$ and $\ket{0}$, just as in Eq.~(\ref{eq: goal})
but by comparing the two states components wise. As such the goal is to find the parameters, $\bar{\phi}$, so that $\ket{\Phi} =\ket{0}$ $\iff$ $U_2(\bar{\phi}) = U^\dagger$. Therefore, we iterate through the parameters $\bar{\phi}$ until the trace distance is minimised to a precision of $C(\bar{\phi})<0.005$. To search the parameters space, we perform a random walk for each phase variable and check whether the goal function is being minimised at each iteration, as shown in Fig. \ref{fig:concept}(d).\clearpage
\section{Experiment}
\label{sec:Experiment}
\subsection{Set-up \& Apparatus}
\label{subsec:Setup}
\begin{figure}[htbp]
    \centering\includegraphics[keepaspectratio, width=1\textwidth]{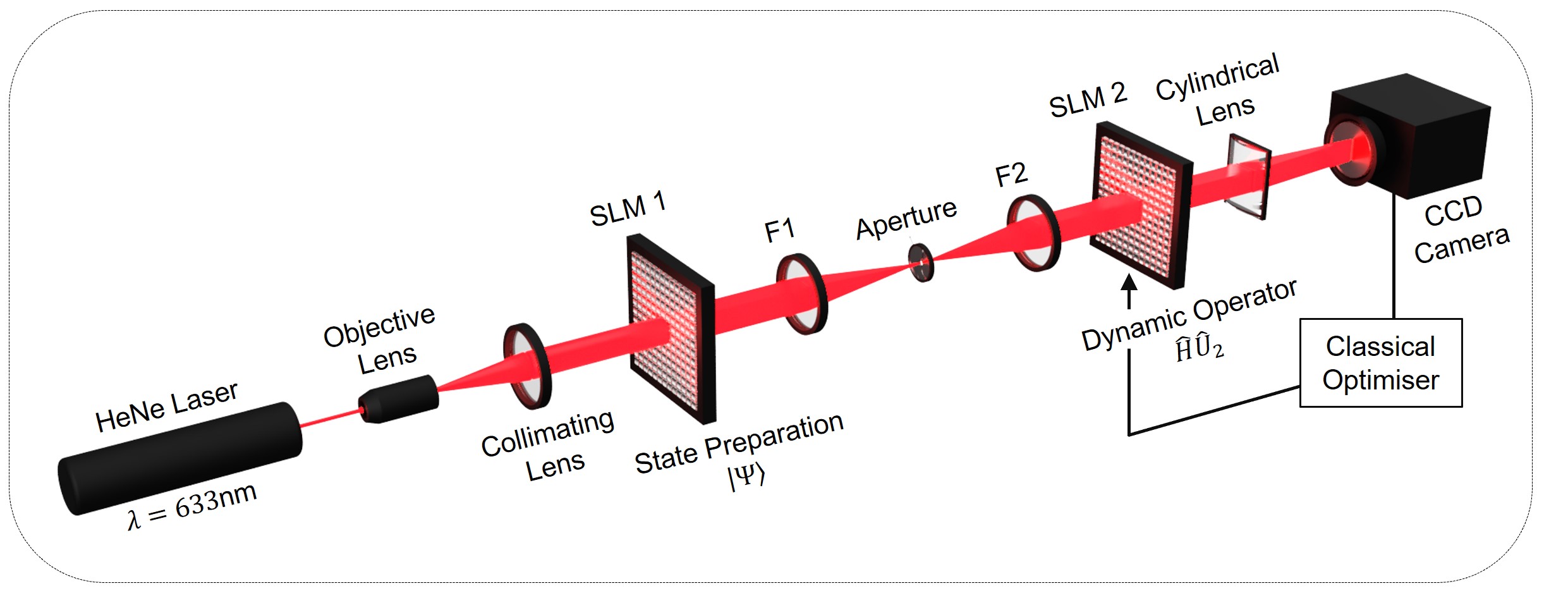}
    \caption{In our experiment, a Helium-Neon (HeNe) laser was expanded by an objective and collimating lens to overfill the first spatial light modulator (SLM 1) on which the initial state vector was prepared, $\ket{\Psi}$.  The resulting optical field was relay imaged with the telescope comprising lenses F1 and F2 to the plane of SLM 2 which encodes the dynamic operator $\hat{H}\hat{U}_2(\bar{\phi})$. The aperture in the Fourier plane of the telescope removed unwanted diffraction orders. The output from the product of the initial state vector and the dynamic operator was passed through a cylindrical lens to perform the necessary row summation, captured by integrating the signal in the $y$-direction on a CCD camera.  Feedback from this output dynamically adjusted the operator on SLM 2 using a classical optimiser until minimisation of the goal function.}
    \label{fig:setup}
\end{figure}
\raggedright
As seen in Fig.~\ref{fig:setup}, we used a $\lambda = 633$ nm wavelength Helium Neon (HeNe) laser which produced a coherent Gaussian beam that was magnified to the first spatial light modulator (SLM 1). Since we magnified and collimated the beam, we ensured that SLM 1 was overfilled with light and that the field was of uniform intensity and therefore approximated a plane wave.  As a result, we could encode information into the transverse  field components  (defined in the $x-y$ plane).\break \\At SLM 1, we prepared the desired input state represented by Eq.~(\ref{eq: psi}), and propagated it to SLM 2. Between the two SLMs, an imaging system composed of two lenses of equal focal length ($f=250$ mm) was used. This also allowed us to place an aperture at the Fourier plane of our imaging system to select only the first diffraction order to be transmitted. On SLM 2, the field  was acted on by the dynamic operator $\hat{M}(\bar{\phi}) = \hat{H} \hat{U}_2(\bar{\phi}) $ following Eq.~(\ref{Eq: statebeforelens}). This resulted in a field where the input state phases were uniquely linked to the phases contained within the dynamic matrix as in Eq.~(\ref{eqn:phaselink}). However, at this point, the field $\ket{\Psi}$ and operator $\hat{M}$ only enacted a point-wise multiplication (or Hadamard product). We then completed the matrix-vector product operation using a cylindrical lens ($f=150$ mm) which performed a one-dimensional Fourier transform - a process that is analogous to a contraction of the rows of the matrix.\break \\The resultant field was captured by a CCD camera and consisted of multiple interference fringes, where the zeroth order fringes were singled out and contained the desired resultant vector, $\ket{\Phi}$, encoded as a vertical strip of light containing the elements $\propto |\Phi_k|^2$ as in Fig.~\ref{fig:concept}(c).  Next, using each of the normalised  measured intensities, we computed the goal function in Eq.~(\ref{eq: goal}) from the normalised measured intensities, $|\Phi_k|^2 1/  \sum_k |\Phi_k|^2$. It is noted that our goal function $C(\bar{\phi})$ is identical to the $\chi ^{2}$-error calculation that is used for maximum likelihood optimisation. At each iteration, the matrix on SLM 2 was altered to represent a new ansatz for, $\hat{H}\hat{U}_2(\bar{\phi})$, which produced a new resultant field that was again compared to the target output. This process was repeated over multiple iterations until the observed output converged to the target output as seen in Fig. \ref{fig:concept}(d). 
\subsection{Results \& Discussion}
\label{subsec:Results}
\subsubsection{State Characterisation}
\label{subsec:statecharacterisation}
\begin{figure}[htbp]
    \centering\includegraphics[keepaspectratio, width=1\textwidth]{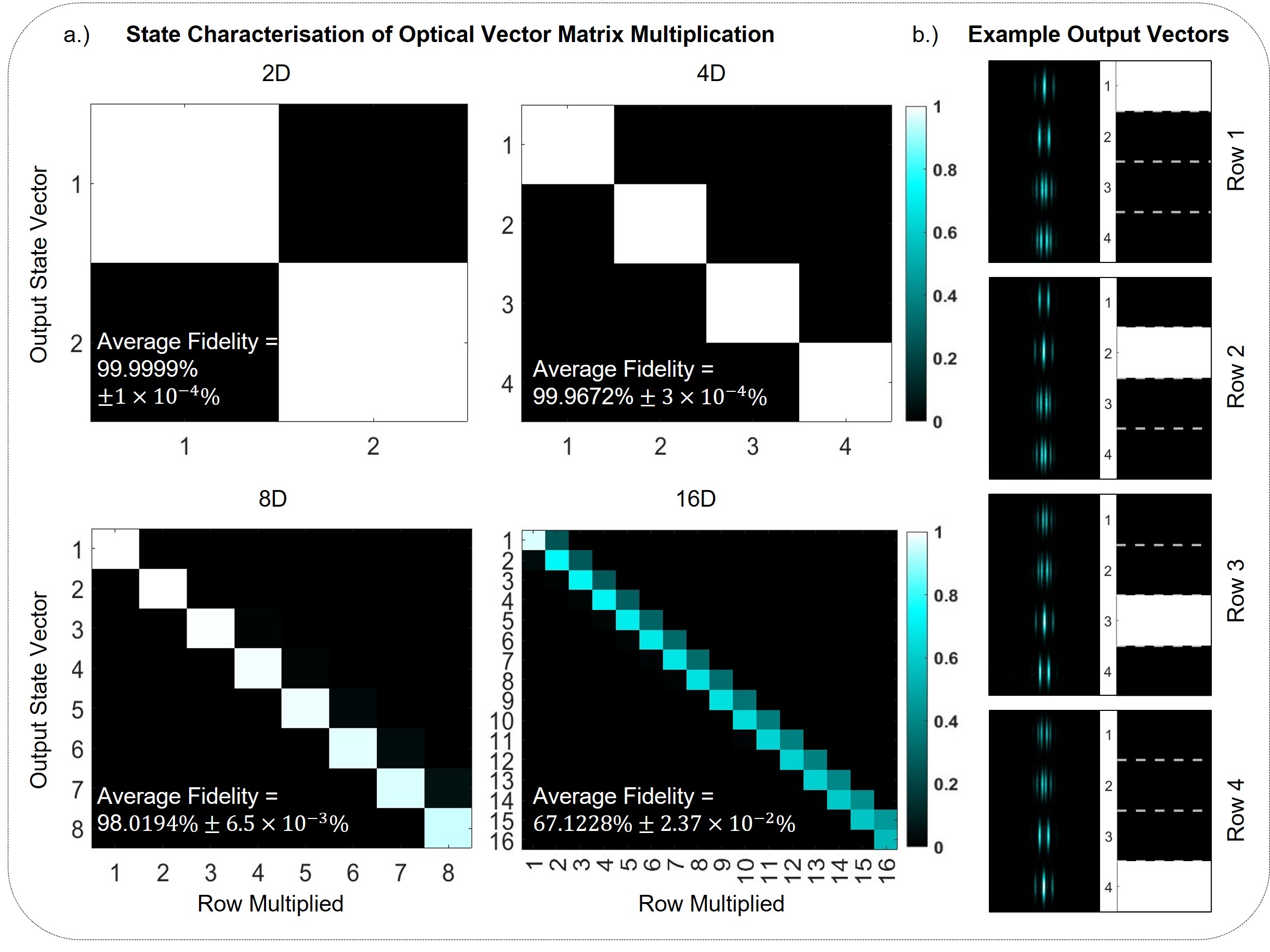}
    \caption{(a) System State Characterisation. Fidelities of the projective measurements from the vector-matrix multiplication for $d$ = 2, 4, 8 and 16. Each column in the respective matrices represents the result of a multiplication between a Hadamard matrix and a vector extracted from the row of the Hadamard matrix which matches the column index. (b) Examples of the individual output fields and measurements that were used to construct the characterisation matrices.}
    \label{fig:characterisation}
\end{figure}
The optical vector-matrix multiplier was first characterised to determine the fidelity of the vector projections. We characterised the system by performing a multiplication between a Hadamard matrix and its various rows - The properties of a Hadamard matrix make it so that the multiplication by one of its rows resulted in a vector with a 1 in the position of the chosen row and 0 in every other row as shown in Fig. \ref{fig:characterisation}(b). This meant that a map of perfect multiplications between a Hadamard matrix and all of its rows resulted in an identity matrix. We therefore compared the results of the multiplications performed by the system to an identity matrix and calculated their fidelity. This characterisation is shown in Fig.~\ref{fig:characterisation}(a), where it is seen that as the dimensions of the matrix where increased, so the elements of the observed vectors became less resolved with a concomitant decrease in fidelity. This was likely due to cross-talk noise from neighbouring pixels in the camera, which made it difficult to differentiate the individual fringe elements when we filtered the output intensity field. This noise can be reduced by increasing the resolution of the camera, by resizing the encoded fields to increase the width of the zeroth order fringe or through further optimisation of the physical experimental setup. The decrease in fidelity with increasing dimension can likely be attributed to a misalignment of optical elements within the physical system, or due to the resolution limits of the CCD camera used to capture the output, since the resolution restricts our ability to resolve the individual interference fringes. Regardless of these difficulties, the search algorithm was observed to be tolerant to random interference's and continued to converge with a high fidelity regardless of noise. Having confirmed that the system itself was sufficiently accurate, we could begin to test the variational unitary algorithm.
\subsubsection{Random Walk Plots}
\label{subsec:randomwalkplots}
\begin{figure}[htbp]
    \centering\includegraphics[keepaspectratio, width=1.05\textwidth]{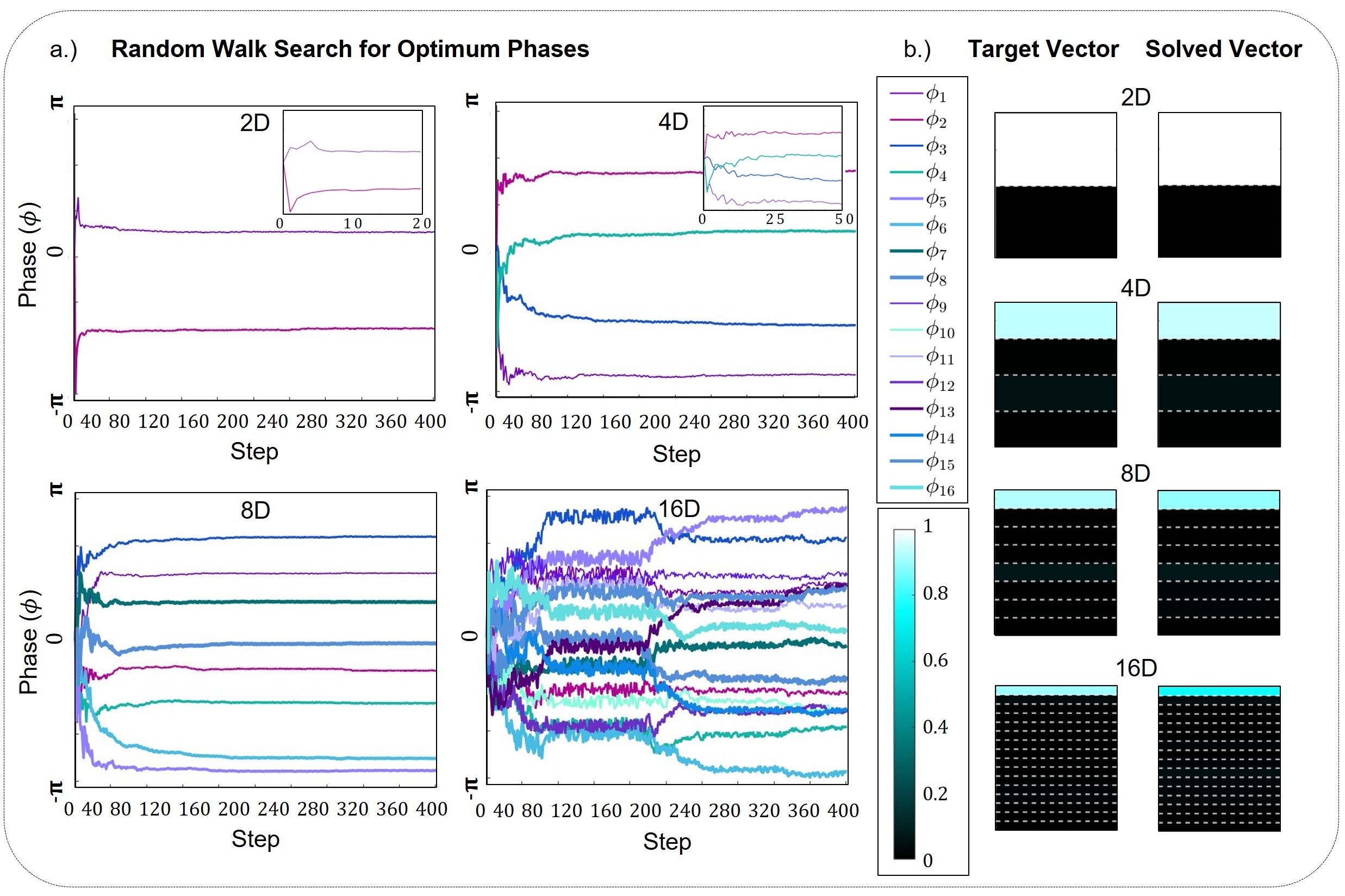}
    \caption{(a) Random walk plots for each tested dimension, showing the independent phase variables, represented by each individual line walk, as they converge to a configuration that minimises a $\chi ^2$ cost function. (b) Examples of the desired output intensity (left) and the average solved output intensity (right).}
    \label{fig:resultswalk}
\end{figure}
The variational approach outlined in Sec. \ref{subsec:Variational Approach} utilised a pseudo random walk method as a search algorithm for the variables $\bar{\phi}$. To assess the performance of the algorithm, the random walk procedure was run 30 times with the maximum length of each random walk being kept constant at 400 steps for control purposes. The boundaries were set at [$-\pi \hspace{0.1cm}\pi$], this allowed for the full $2\pi$ phase range to be searched for a solution. The step size of each walk at each step was calculated using the $\chi^{2}$-error (which is analogous to $C(\bar{\phi})$) between the observed and target output fields, therefore, as the phases of the unitary matrix converged to their desired values, the step size decreased in parallel. Examples of these random walk plots are shown in Fig. \ref{fig:resultswalk}(a) - it is seen that the 2D and 4D plot converge quickly and therefore the first few steps of each plot have been enlarged for clarity. Representation of output state characterisations for each tested dimension (2D, 4D, 8D and 16D) are shown in Fig. \ref{fig:resultswalk}(b) - these examples were chosen to represent the average of the optimum output states across all test runs. They were obtained using the process outlined in Fig. \ref{fig:concept} and we see that regardless of the noise at higher dimensions, the variational algorithm was still able to converge towards a high-fidelity solution. 
\subsubsection{Fidelity Measurements}
\label{subsec:fidelity}
\begin{figure}[htbp]
    \centering\includegraphics[keepaspectratio, width=1\textwidth]{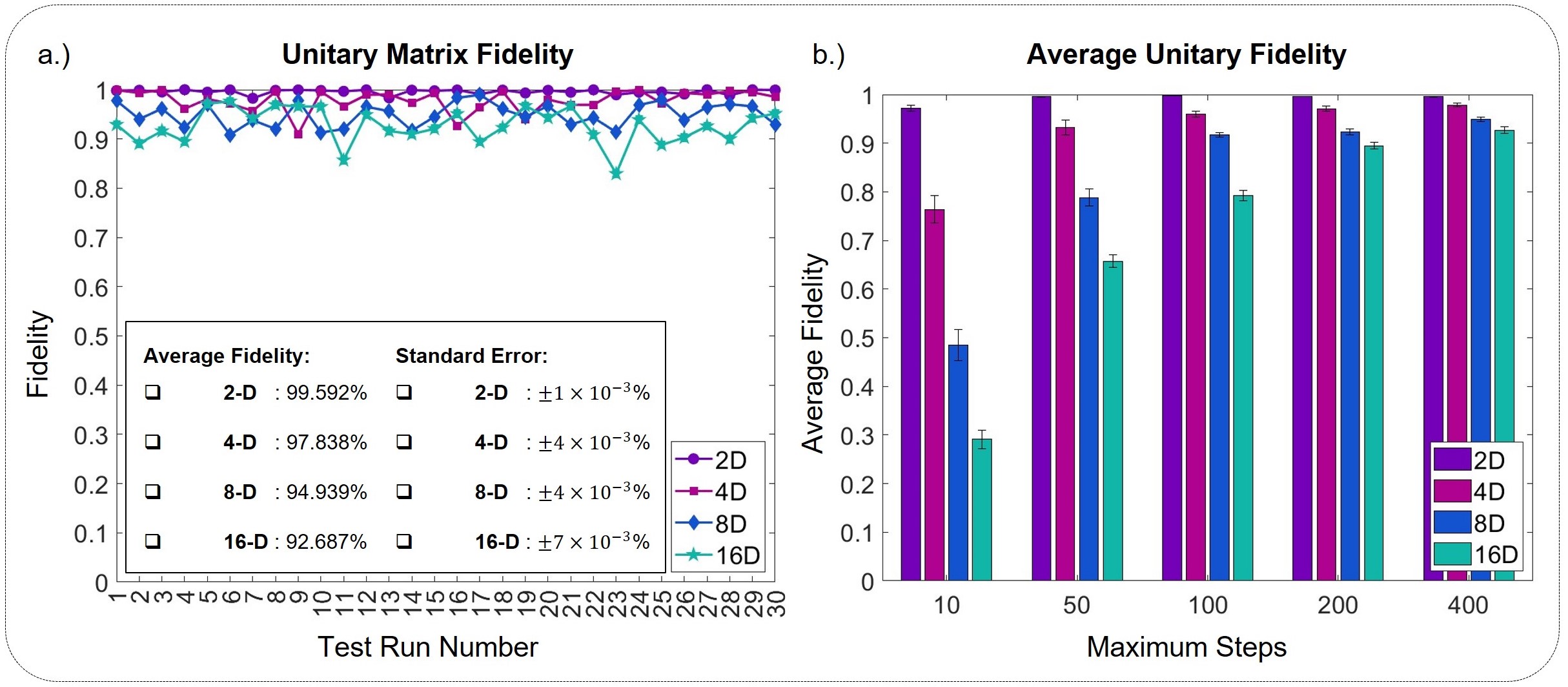}
    \caption{(a) Fidelity measurement for each run for 2, 4, 8 and 16 dimensions with the error bars showcasing the standard error for each dimension, however, the error bars are smaller than the data points. (b) The average fidelity for each dimension as the maximum number of steps in each run is increased, where the error bars represent the standard error as before.}
    \label{fig:resultsfid}
\end{figure}
To determine the viability of the experimentally solved solution, we calculate its fidelity against a target unitary. The target unitary chosen was a diagonal matrix similar to those in Eq.~(\ref{eqn:diagonalunitary}). The phases in the target unitary were the input state phases themselves, the Hermitian conjugate of this target unitary would be the optimum unitary for canceling the phases of the input state. The fidelity of the experimentally solved phases were calculated as
\begin{equation}
    \label{eqn:fidelity}
    F = \bigg| \frac{\text{Tr}(\hat{U}^{\dag}_\text{solved}\cdot \hat{U}_\text{target})}{d} \bigg|,
\end{equation}
where $\text{Tr}$ is the trace. We take the trace of the product between the experimentally solved unitary and the target unitary - if the phases of the solved unitary matches the phases in the expected unitary sufficiently, the product between the two will result in a identity matrix that's trace will be equal to the length of the unitary - this result is therefore normalised by dividing by the dimension (d) of the vector that is being solved for. Taking the absolute value of this results in a value that lies in the range [$0,1$] where a higher fidelity corresponds to a stronger correlation between the solved and target unitary's. The fidelities for each dimension was calculated for $N=30$ runs, shown in Fig. \ref{fig:resultsfid}(a) - with the standard error in these fidelities being calculated as, $\text{SE} = \frac{\sigma}{\sqrt{N}}$,
where $\sigma$ is the sample standard deviation of the fidelities from all the test runs and $N$ is the number of runs. On average, the experimental unitary matrices were found with a fidelity of $F = 99.592\% \pm 1\times10^{-3}\%$ for $d = 2$, $F = 97.838\% \pm 4\times10^{-3}\%$ for $d = 4$, $F = 94.939\% \pm 4\times10^{-3}\%$ for $d = 8$ and $F = 92.687\% \pm 7\times10^{-3}\%$ for $d = 16$. It can be seen that the $d = 2$ case consistently achieved fidelities greater than $99\%$, the $d = 4$ fidelities remained greater than $95\%$ and the $d = 8$ and $d = 16$ cases oscillated comfortably above $90\%$ fidelity. These high fidelity values suggest that the system successfully solved for a unitary matrix that both cancels the phases applied to the input beam and sufficiently matches the phases contained in target unitary. However, it must be noted that unitary matrices are not unique by nature, therefore, the phases of the solved unitary matrix may differ from the input phases by a global phase factor.\break \\  The maximum number of steps for the random walk was varied and the average fidelities and their relative standard errors were recorded in Fig. \ref{fig:resultsfid}(b) - it is clear that as the maximum number of steps increases so does the average fidelity, whilst the standard error decreases. This infers that the accuracy of the result will improve if it is given more time to converge, meaning that lower fidelities at higher dimensions is not necessarily observed due to a limitation in the system. Rather, it is more likely due to the increased number of variables that are needing to be solved for, therefore, more iterations are needed to minimise the global error between the various phases elements.
\section{Discussion}
As presented, the results show that our technique, which is inspired by variational computing approaches, but implemented using an optical vector-matrix multiplier,  is both robust to noise and capable of learning unitary matrices by minimising a goal function that makes use of projective measurements of a resultant field. This approach can be applied to the reconstruction of various channels, matrices of complex and disordered mediums \cite{valencia2020unscrambling, popoff2010measuring} and perhaps for simulations of simple molecules, as was initially intended by the developers of variational quantum computing framework  \cite{peruzzo2014variational}.  This is thanks to the re-configurable nature of SLMs and the and matrix multiplication framework \cite{spall2020fully}. The re-configurable nature of SLMs makes our approach dynamic and adaptable.  Besides SLMs, re-configurable metasurfaces are also a promising candidate as they allow for sub-wavelength modulation of optical fields, thereby increasing encoding dimensions and resolution \cite{zhang2020optically, he2019tunable}. This can help overcome challenges with scaling the system up; since the number of variables to be solved increases the resolution requirements of the system.\break \\Furthermore, another challenge to overcome as the dimensions of the matrices are increased is the number of iterations before convergence, as this can increase the convergence time of the approach. Besides using optimal strategies, such as gradient based parameter selection,  increasing the modulation and detection speed of our dynamic unitary (SLM) and camera, respectively, can compensate for the increased time complexity due to the parameters \cite{parsopoulos2002recent}. In traditional VQAs,  most algorithms are pared with a classical optimiser, e.g. genetic algorithm \cite{lambora2019genetic} or particle swarm  algorithm\cite{parsopoulos2002recent} \cite{cerezo2021variational}. Adapting our random walk approach with such optimisation techniques can help.\break \\On the other hand, faster modulation devises such as digital micro-mirror devices (DMDs) \cite{ren2015tailoring, mirhosseini2013rapid, scholes2020structured} can be implemented, potentially reaching modulation frequencies in the kHz regime (with others promising modulation speeds in the MHz regime \cite{ye2021high}). Furthermore, since the array of values that determine the output results  (measurement outcomes) requires single pixel measurements, high speed CMOS cameras or SPAD arrays can be adapted to match the modulation speed of the modulation device.\clearpage
\section{Conclusion}
\label{sec:Conclusion}
We have demonstrated a technique that blends optical matrix multiplication and variational quantum computing to learn unitary transmission matrices in photonics systems where the transverse spatial degree of freedom is used as a resource. We showed that the problem of learning a unitary operation that acts as a perturbation on a laser field, given a suitable goal function, is similar to finding the ground state energy of a Hamiltonian given  a corresponding parameterised unitary that can be used to search the parameter space. In our approach we used a lattice/grid of Gaussian modes as our encoding basis, where the transverse Cartesian coordinates act as two registers that can be controlled in-order to enact operations - which is useful for extracting measurements that can be used to compute the goal/cost function (analogous to the observable of the problem Hamiltonian).  The dynamic nature of our approach makes the scheme diverse and adaptable and we showcased this approach using 2-dimensional to 16-dimensional encoding.
\section{Acknowledgments}
\label{sec:Acknowledgments}
The authors would like to acknowledge the support of Optica for funding the Emerging Leader in Optics Chair, South African Quantum initiative (SAQuti), Department of Science and Innovation (South Africa), and the National Research Foundation (South Africa).


\begin{thebibliography}{10}
\newcommand{\enquote}[1]{``#1''}

\bibitem{forbes2021structured}
A.~Forbes, M.~de~Oliveira, and M.~R. Dennis, \enquote{Structured light,} {\protect\JournalTitle{Nature Photonics}} \textbf{15}, 253--262 (2021).

\bibitem{nape2023quantum}
I.~Nape, B.~Sephton, P.~Ornelas, \emph{et~al.}, \enquote{Quantum structured light in high dimensions,} {\protect\JournalTitle{APL Photonics}} \textbf{8} (2023).

\bibitem{yao2011orbital}
A.~M. Yao and M.~J. Padgett, \enquote{Orbital angular momentum: origins, behavior and applications,} {\protect\JournalTitle{Advances in optics and photonics}} \textbf{3}, 161--204 (2011).

\bibitem{da2021path}
B.~Da~Lio, D.~Cozzolino, N.~Biagi, \emph{et~al.}, \enquote{Path-encoded high-dimensional quantum communication over a 2-km multicore fiber,} {\protect\JournalTitle{npj Quantum Information}} \textbf{7}, 63 (2021).

\bibitem{valencia2020unscrambling}
N.~H. Valencia, S.~Goel, W.~McCutcheon, \emph{et~al.}, \enquote{Unscrambling entanglement through a complex medium,} {\protect\JournalTitle{Nature Physics}} \textbf{16}, 1112--1116 (2020).

\bibitem{valencia2020high}
N.~H. Valencia, V.~Srivastav, M.~Pivoluska, \emph{et~al.}, \enquote{High-dimensional pixel entanglement: efficient generation and certification,} {\protect\JournalTitle{Quantum}} \textbf{4}, 376 (2020).

\bibitem{wang2022orbital}
J.~Wang, J.~Liu, S.~Li, \emph{et~al.}, \enquote{Orbital angular momentum and beyond in free-space optical communications,} {\protect\JournalTitle{Nanophotonics}} \textbf{11}, 645--680 (2022).

\bibitem{huang2014100}
H.~Huang, G.~Xie, Y.~Yan, \emph{et~al.}, \enquote{100 tbit/s free-space data link enabled by three-dimensional multiplexing of orbital angular momentum, polarization, and wavelength,} {\protect\JournalTitle{Optics letters}} \textbf{39}, 197--200 (2014).

\bibitem{cozzolino2019high}
D.~Cozzolino, B.~Da~Lio, D.~Bacco, and L.~K. Oxenl{\o}we, \enquote{High-dimensional quantum communication: benefits, progress, and future challenges,} {\protect\JournalTitle{Advanced Quantum Technologies}} \textbf{2}, 1900038 (2019).

\bibitem{mirhosseini2015high}
M.~Mirhosseini, O.~S. Maga{\~n}a-Loaiza, M.~N. O’Sullivan, \emph{et~al.}, \enquote{High-dimensional quantum cryptography with twisted light,} {\protect\JournalTitle{New Journal of Physics}} \textbf{17}, 033033 (2015).

\bibitem{sit2017high}
A.~Sit, F.~Bouchard, R.~Fickler, \emph{et~al.}, \enquote{High-dimensional intracity quantum cryptography with structured photons,} {\protect\JournalTitle{Optica}} \textbf{4}, 1006--1010 (2017).

\bibitem{shikder2022image}
A.~Shikder, P.~Kumar, and N.~K. Nishchal, \enquote{Image encryption by structured phase encoding and its effectiveness in turbulent medium,} {\protect\JournalTitle{IEEE Photonics Technology Letters}} \textbf{35}, 128--131 (2022).

\bibitem{wang2021high}
Q.-K. Wang, F.-X. Wang, J.~Liu, \emph{et~al.}, \enquote{High-dimensional quantum cryptography with hybrid orbital-angular-momentum states through 25 km of ring-core fiber: A proof-of-concept demonstration,} {\protect\JournalTitle{Physical Review Applied}} \textbf{15}, 064034 (2021).

\bibitem{sun20133d}
B.~Sun, M.~P. Edgar, R.~Bowman, \emph{et~al.}, \enquote{3d computational imaging with single-pixel detectors,} {\protect\JournalTitle{Science}} \textbf{340}, 844--847 (2013).

\bibitem{sephton2023revealing}
B.~Sephton, I.~Nape, C.~Moodley, \emph{et~al.}, \enquote{Revealing the embedded phase in single-pixel quantum ghost imaging,} {\protect\JournalTitle{Optica}} \textbf{10}, 286--291 (2023).

\bibitem{moodley2023all}
C.~Moodley and A.~Forbes, \enquote{All-digital quantum ghost imaging: tutorial,} {\protect\JournalTitle{JOSA B}} \textbf{40}, 3073--3095 (2023).

\bibitem{cheng2020simulate}
K.~Cheng, W.~Zhang, Z.~Wei, \emph{et~al.}, \enquote{Simulate deutsch-jozsa algorithm with metamaterials,} {\protect\JournalTitle{Optics Express}} \textbf{28}, 16230--16243 (2020).

\bibitem{perez2018first}
B.~Perez-Garcia, R.~I. Hernandez-Aranda, A.~Forbes, and T.~Konrad, \enquote{The first iteration of grover's algorithm using classical light with orbital angular momentum,} {\protect\JournalTitle{Journal of Modern Optics}} \textbf{65}, 1942--1948 (2018).

\bibitem{kumar2022optical}
P.~Kumar, N.~K. Nishchal, T.~Omatsu, and A.~S. Rao, \enquote{Optical vortex array for two-dimensional exclusive-or operation,} {\protect\JournalTitle{Applied Physics B}} \textbf{128}, 98 (2022).

\bibitem{dharnidharka2021optical}
M.~Dharnidharka, U.~Chadha, L.~M. Dasari, \emph{et~al.}, \enquote{Optical tomography in additive manufacturing: a review, processes, open problems, and new opportunities,} {\protect\JournalTitle{The European Physical Journal Plus}} \textbf{136}, 1133 (2021).

\bibitem{mohseni2008quantum}
M.~Mohseni, A.~T. Rezakhani, and D.~A. Lidar, \enquote{Quantum-process tomography: Resource analysis of different strategies,} {\protect\JournalTitle{Physical Review A}} \textbf{77}, 032322 (2008).

\bibitem{endo2021hybrid}
S.~Endo, Z.~Cai, S.~C. Benjamin, and X.~Yuan, \enquote{Hybrid quantum-classical algorithms and quantum error mitigation,} {\protect\JournalTitle{Journal of the Physical Society of Japan}} \textbf{90}, 032001 (2021).

\bibitem{cerezo2021variational}
M.~Cerezo, A.~Arrasmith, R.~Babbush, \emph{et~al.}, \enquote{Variational quantum algorithms,} {\protect\JournalTitle{Nature Reviews Physics}} \textbf{3}, 625--644 (2021).

\bibitem{peruzzo2014variational}
A.~Peruzzo, J.~McClean, P.~Shadbolt, \emph{et~al.}, \enquote{A variational eigenvalue solver on a photonic quantum processor,} {\protect\JournalTitle{Nature communications}} \textbf{5}, 4213 (2014).

\bibitem{xu2021variational}
X.~Xu, J.~Sun, S.~Endo, \emph{et~al.}, \enquote{Variational algorithms for linear algebra,} {\protect\JournalTitle{Science Bulletin}} \textbf{66}, 2181--2188 (2021).

\bibitem{bravo2023variational}
C.~Bravo-Prieto, R.~LaRose, M.~Cerezo, \emph{et~al.}, \enquote{Variational quantum linear solver,} {\protect\JournalTitle{Quantum}} \textbf{7}, 1188 (2023).

\bibitem{atchade2022quantum}
P.~Atchade-Adelomou, \enquote{Quantum algorithms for solving hard constrained optimisation problems,} {\protect\JournalTitle{arXiv preprint arXiv:2202.13125}}  (2022).

\bibitem{martin2022quantum}
J.~D. Mart{\'\i}n-Guerrero and L.~Lamata, \enquote{Quantum machine learning: A tutorial,} {\protect\JournalTitle{Neurocomputing}} \textbf{470}, 457--461 (2022).

\bibitem{forbes2016creation}
A.~Forbes, A.~Dudley, and M.~McLaren, \enquote{Creation and detection of optical modes with spatial light modulators,} {\protect\JournalTitle{Advances in Optics and Photonics}} \textbf{8}, 200--227 (2016).

\bibitem{ayoub2021high}
A.~B. Ayoub and D.~Psaltis, \enquote{High speed, complex wavefront shaping using the digital micro-mirror device,} {\protect\JournalTitle{Scientific Reports}} \textbf{11}, 18837 (2021).

\bibitem{zhang2020optically}
X.~G. Zhang, W.~X. Jiang, H.~L. Jiang, \emph{et~al.}, \enquote{An optically driven digital metasurface for programming electromagnetic functions,} {\protect\JournalTitle{Nature Electronics}} \textbf{3}, 165--171 (2020).

\bibitem{spall2020fully}
J.~Spall, X.~Guo, T.~D. Barrett, and A.~Lvovsky, \enquote{Fully reconfigurable coherent optical vector--matrix multiplication,} {\protect\JournalTitle{Optics Letters}} \textbf{45}, 5752--5755 (2020).

\bibitem{ndagano2017characterizing}
B.~Ndagano, B.~Perez-Garcia, F.~S. Roux, \emph{et~al.}, \enquote{Characterizing quantum channels with non-separable states of classical light,} {\protect\JournalTitle{Nature Physics}} \textbf{13}, 397--402 (2017).

\bibitem{cerezo2021cost}
M.~Cerezo, A.~Sone, T.~Volkoff, \emph{et~al.}, \enquote{Cost function dependent barren plateaus in shallow parametrized quantum circuits,} {\protect\JournalTitle{Nature communications}} \textbf{12}, 1791 (2021).

\bibitem{popoff2010measuring}
S.~Popoff, G.~Lerosey, R.~Carminati, \emph{et~al.}, \enquote{Measuring the transmission matrix in optics: an approach to the study and control of light propagation in disordered media,} {\protect\JournalTitle{Physical review letters}} \textbf{104}, 100601 (2010).

\bibitem{he2019tunable}
Q.~He, S.~Sun, and L.~Zhou, \enquote{Tunable/reconfigurable metasurfaces: physics and applications,} {\protect\JournalTitle{Research}}  (2019).

\bibitem{parsopoulos2002recent}
K.~E. Parsopoulos and M.~N. Vrahatis, \enquote{Recent approaches to global optimization problems through particle swarm optimization,} {\protect\JournalTitle{Natural computing}} \textbf{1}, 235--306 (2002).

\bibitem{lambora2019genetic}
A.~Lambora, K.~Gupta, and K.~Chopra, \enquote{Genetic algorithm-a literature review,} in \emph{2019 international conference on machine learning, big data, cloud and parallel computing (COMITCon),}  (IEEE, 2019), pp. 380--384.

\bibitem{ren2015tailoring}
Y.-X. Ren, R.-D. Lu, and L.~Gong, \enquote{Tailoring light with a digital micromirror device,} {\protect\JournalTitle{Annalen der physik}} \textbf{527}, 447--470 (2015).

\bibitem{mirhosseini2013rapid}
M.~Mirhosseini, O.~S. Magana-Loaiza, C.~Chen, \emph{et~al.}, \enquote{Rapid generation of light beams carrying orbital angular momentum,} {\protect\JournalTitle{Optics express}} \textbf{21}, 30196--30203 (2013).

\bibitem{scholes2020structured}
S.~Scholes, R.~Kara, J.~Pinnell, \emph{et~al.}, \enquote{Structured light with digital micromirror devices: a guide to best practice,} {\protect\JournalTitle{Optical Engineering}} \textbf{59}, 041202--041202 (2020).

\bibitem{ye2021high}
X.~Ye, F.~Ni, H.~Li, \emph{et~al.}, \enquote{High-speed programmable lithium niobate thin film spatial light modulator,} {\protect\JournalTitle{Optics Letters}} \textbf{46}, 1037--1040 (2021).

\end{thebibliography}

\end{document}